\documentclass[12pt,english,aps,manuscript]{article}
\usepackage[latin9]{inputenc}
\usepackage{float}

\makeatletter

\providecommand{\tabularnewline}{\\}

\newcommand{\lyxaddress}[1]{
\par {\raggedright #1
\vspace{1.4em}
\noindent\par}
}

\makeatother

\usepackage{babel}

\begin{document}

\title{Analytic Calculation of Neutrino Mass Eigenvalues }

\author{K.Chaturvedi,Bipin Singh Koranga and Vinod Kumar}

\maketitle

\lyxaddress{Department of Physics, Bundelkhand University, Jhansi-284128, India}

\lyxaddress{Department of Physics, Kirori Mal college (University of Delhi,)
Delhi-110007, India}

\lyxaddress{Department of Physics , University of Delhi, Delhi-110007, India}
\begin{abstract}
Implicaion of the neutrino oscillation search for the neutrino mass
square difference and mixing are discussed. We have considered the
effective majorana mass $m_{ee}$, related for $\beta\beta_{0\nu}$decay.
We find limits for neutrino mass eigen value $m_{i}$ in the different
neutrino mass spectrum,which explain the different neutrino data.
\end{abstract}

\section{Introduction}

The phenomenon of neutrino oscillation, impressive advance have been
made to understand the phenomenology of neutrino oscillation through
solar neutrino,atmospheric neutrino,reactor neutrino and accelerator
neutrino experiment. The experimental research on the nature of neutrinos
through terrestrial as well as extra-terrestrial approaches has finally
confirmed the neutrino oscillation in atmospheric {[}1,2,3,4{]} solar
{[}5,6,7,8,9{]}, reactor {[}10, 11,12{]} and accelerator {[}13,14{]}
neutrino sources, establishing that neutrinos have mass. Furthermore,
it is generally agreed that oscillations among three neutrino species
are sufficient to explain the atmospheric, solar, reactor and accelerator
neutrino puzzle. The neutrino oscillation experiments provide us with
neutrino mass square differences, mixing angles and a possible hierarchy
in the neutrino mass spectrum. The main physical goal in future experiment
are the determination of the unknown parameter $\theta_{13}$ . In
particular, the observation of $\delta$ is quite interesting for
the point of view that~$\delta$ related to the origin of the matter
in the universe. One of the most important parameter in neutrino physics
is the magnitude of mixing angle $\theta_{13}$ and CP phase $\delta$.
The oscillation data also sugges that the neutrinos may belong to
either a normal hierarchy ($m_{1}<m_{2}<m_{3})$ or an inverted hierarchy
($m_{3}<m_{1}<m_{2}$). The data do not exclude the possibility that
the mass of the light neutrino could be much larger than $\sqrt{\bigtriangleup_{31}}$
, which would imply the possible existence of a quasi-degenerate neutrino
mass spectrum ($m_{1}\approx m_{2}\approx m_{3}$). On the other hand,
the actual mass of neutrinos cannot be extracted from these data,
only the study of tritium single $\beta$ decay and nuclear neutrino-less
double beta decay together can provide sharpest limits on the mass
and nature of neutrinos. Neutrino oscillations, which only depend
on mass square difference, give no information about the absolute
value of the neutrino mass squared eigenvalues . Hence, there are
various possibilities of neutrino hierarchy spectrums consistent with
solar and atmospheric neutrino oscillation data. The mass eigenstates
with eigen values $m_{i}$ can be determined if the absolute value
of effective mass of neutrino is exactly known. The current neutrino-less
double beta decay experiments only provide the upper limit on effective
Majorana neutrino mass $<m_{ee}>$ so that absolute scale of neutrino
mass is not determined yet. Therefore, in the present work we have
attempted to present a picture of neutrino mass spectrum in the case
of normal, inverted and almost degenerate hierarchy of neutrino masses
by taking some specific choices of effective mass. 

In this paper, we will discuss the masses of the vacuum eigenstates
$m_{1},$$m_{2}$ and $m_{3}$for different neutrino mass spectrum,
namely normal mass hierarchy ($m_{1}<m_{2}<m_{3}),$inverted mass
hierarchy ($\ensuremath{m_{3}<m_{1}<m_{2})}$ and almost degenerate
spectrum $(m_{1}\approx m_{2}\approx m_{3}$).The present work is
organized as follows.In Sec.2, we outline the neutrino oscillation
parameters. In Sec.3, we have given the theoretical formalism to calculate
mass eigenvalues $m_{i}$ for above mentioned hierarchies. In Sec.4,
we present the numerical results and Sec.5 is devoted to the conclusions.

\section{Mixing Angles and Neutrino Mass Squared Differences}

The first evidence is the observation of zenith-angle dependence of
atmospheric neutrino defect {[}15{]} dependent of the atmospheric
neutrino $\nu_{\mu}\rightarrow\nu_{\mu}$ transition with the mass
difference and the mixing as 

\begin{equation}
\Delta_{31}=(1-2)\times10^{-3}eV^{2},sin^{2}2\theta_{23}=1.0.\end{equation}

The second evidence is the solar neutrino deficit {[}16{]}, which
is consistent with $\nu_{\mu}\rightarrow\nu_{\tau}/\nu_{e}$ transition.
The SNO experiments {[}17{]} are consistent with the standard solar
model {[}18{]} and strongly suggest the LMA solution. 

\begin{equation}
\Delta_{21}=7\times10^{-5}eV^{2},sin^{2}2\theta_{12}=0.8.\end{equation}

Solar neutrino experiments (Super-K, GALLEX, SAGE, SNO and GNO) show
the neutrino oscillations, neutrino oscillation provide the most elegant
explanation of all the data {[}19{]}.

\begin{equation}
\Delta_{solar}=7_{-1.3}^{+5}\times10^{-5}eV^{2},\end{equation}

\begin{equation}
tan^{2}\theta_{solar}=0.4{}_{-0.1}^{+0.14}.\end{equation}

Atmospheric neutrino experiments ( Kamiokande, Super-K ) also show
the neutrino oscillation. The most excellent fit to the all data {[}19{]}.

\begin{equation}
\Delta_{atmo}=2.0_{-0.92}^{+1.0}\times10^{-3}eV^{2},\end{equation}

\begin{equation}
sin^{2}2\theta_{atmo}=0.4{}_{-0.10}^{+0.14}.\end{equation}

The CHOOZ reactor experiment {[}20{]} gives the upper bound of the
third mixing angle $\theta_{13}$as 

\begin{equation}
sin^{2}\theta_{13}<0.20\,\,\,\,\,\,\,\,\,\,\,\,\,\,\,\,\,\: for\,\,\,\,\,\,\,\,\,\,\,\,\,\,\,\,\,\,\,\,\,\,\,\,\,\,\,\,|\Delta_{31}|=2.0\times10^{-3}eV^{2},\end{equation}

\begin{equation}
sin^{2}\theta_{13}<0.16\,\,\,\,\,\,\,\,\,\,\,\,\,\,\,\,\,\: for\,\,\,\,\,\,\,\,\,\,\,\,\,\,\,\,\,\,\,\,\,\,\,\,\,\,\,\,|\Delta_{31}|=2.5\times10^{-3}eV^{2},,\end{equation}

\begin{equation}
sin^{2}\theta_{13}<0.14\,\,\,\,\,\,\,\,\,\,\,\,\,\,\,\,\,\: for\,\,\,\,\,\,\,\,\,\,\,\,\,\,\,\,\,\,\,\,\,\,\,\,\,\,\,\,|\Delta_{31}|=3.0\times10^{-3}eV^{2},\end{equation}

at the 90 \% CL. The CP phase $\delta$ has not been constrained.
The future neutrino experiments plan to measure the oscillation parameters
precisely.

\section{Effective Majorona Mass of Electron Neutrino}

In the presence of three flavour neutrino mixing the electron neutrino
is combination of mass eigenstate, $\nu_{i}$ with eigenvalue $m_{i}$

\begin{equation}
\nu_{e}=\sum U_{ej}\nu_{ij}\,\,\, i=1,2,3.\end{equation}

Here $U_{ei}$are the elements of the mixing matrix,which relates
the flavour states the the mass eigenstates. The $\beta\beta_{0\nu}$
decay rate is determined by the effective Majorana mass of the electron
neutrino $m_{ee}$. Under the assumption of three flavour neutrino
mixing of neutrino, the effective Majorana neutrino mass $m_{ee}$
is

\[
|m_{ee}|=|\sum|U_{ej}|^{2}e^{i\phi_{j}}m_{j}|\]

\begin{equation}
=|c_{13}^{2}c_{12}^{2}m_{1}+c_{13}^{2}s_{12}^{2}e^{i\phi_{2}}m_{2}+s_{13}^{2}e^{i\phi_{3}}m_{3}|.\end{equation}

Where 

$|U_{ej}|,$j=1,2,3 are the absolute values of the elements of the
first row of neutrino mixing matrix,

\[
U=\left(\begin{array}{ccc}
c_{12}c_{13} & s_{12}c_{13} & s_{13}e^{-i\delta}\\
-s_{12}c_{23}-c_{12}s_{23}s_{13}e^{i\delta} & c_{12}c_{23}-s_{12}s_{23}s_{13}e^{i\delta} & s_{23}c_{13}\\
s_{12}s_{23}-c_{12}c_{23}s_{13}e^{i\delta} & -c_{12}s_{23-}s_{12}s_{13}s_{23}e^{i\delta} & c_{23}s_{13}\end{array}\right),\]

here $c_{ij}=cos\theta_{ij},$$s_{ij}=sin\theta_{ij}$. The angle
$\phi_{2},$$\phi_{3}$ are the two majorana CP phase. The masses
of the vacuum eigenstates are takes to be $m_{1},$$m_{2}$ and $m_{3}.$
On squaring eq(11),

\begin{equation}
|m_{ee}|^{2}=A^{2}m_{1}^{2}+\mathbf{B}^{2}m_{2}^{2}+C^{2}m_{3}^{2}+2ABm_{1}m_{2}cos\phi_{2}+2ACm_{1}m_{3}cos\phi_{3}+2BCm_{3}m_{2}cos(\phi_{2}-\phi_{3}),\end{equation}

where $A=c_{3}^{2}c_{2}^{2}$,$B=c_{3}^{2}s_{2}^{2}$, $C=s_{3}^{2}$

\section{Numerical Results}

In table-1,table-3 and table-3, we list neutrino mass eigenvalue $m_{i}$
in case of different neutrino mass spectrum, using the best fit value
given in ref{[}21{]}.We have varied $m_{ee}$ from 1- 10 meV and 10-100
meV in case of normal and inverted hierarchy respectively. For degenerate
case,$m_{1}\approx m_{2}\approx m_{3}$, and considering the analysis
of the Sloan Digital Survey Data and WMAP data {[}22{]}, Wilkinson
Microwave Anisotropy Probe (WMAP) and 2 degree Field Galaxy Redshift
Survey (2dFGRS) data {[}23{]}, we have varied $\mbox{\ensuremath{m_{ee}}}$
from 100-600 meV. For normal hierarchy, $m_{1}$ varies from 0.0002eV
- 0.01eV and $m_{2}$ varies from 0.009 eV- 0.013eV, while variation
in $m_{3}$ is from 0.049-0.056 eV. The variation in $m_{3}$ is quite
less in comparison to $m_{2}$ and $m_{1}$. In case of inverted hierarchy,
$m_{1}$ and $m_{2}$ both vary from 0.01-0.263 eV while $m_{3}$
varies from 0.047-0.259 eV for considered values of $\mbox{\ensuremath{m_{ee}}}$.
For 10-30 meV range of $m_{ee}$, $m_{3}$ values are higher than
$m_{1}$ and $m_{2}$, while for 40-100 meV range of $\mbox{\ensuremath{m_{ee}}}$,
$m_{3}$ value is lower than $m_{1}$ and $m_{2}$. The almost same
variation for $m_{1}$ and $m_{2}$ also confirms the consideration
of $m_{1}$, $m_{2}$, for inverted hierarchy. In case of AD hierarchy,
all the $m_{1}$, $m_{2}$, $m_{3}$ values lie  in the range from
0.26-1.74 eV. We compute the neutrino mass eigenvalue using eq (11).
We have taken normal mass hierarchy $\Delta_{31}>0$ , inverted mass
hierarchy$\Delta_{31}<0$ and almost degenerated case. For simplicity,
we have set the majorana phases $\phi=0^{o},180^{o}.$ 

\begin{table}[H]
\begin{tabular}{|c|c|c|c|c|}
\hline 
Mass Hierarchy & $m_{\nu}$ & Mass Eigenstate & Majorana Phases & Majorana Phases\tabularnewline
\hline
\hline 
 & (eV) & (eV) & $\phi_{2}=0^{0},\phi_{3}=180^{0}$ & $\phi_{3}=0^{0},\phi_{2}=180^{0}$\tabularnewline
\hline 
Normal & 0.010 & $m_{1}$ & $1.0\times10^{-2}$ & $2.62\times10^{-2}$\tabularnewline
\hline 
 &  & $m_{2}$ & $1.33\times10^{-2}$ & $2.77\times10^{-2}$\tabularnewline
\hline 
 &  & $m_{3}$ & $5.05\times10^{-2}$ & $5.60\times10^{-2}$\tabularnewline
\hline 
 & 0.009 & $m_{1}$ & $8.89\times10^{-3}$ & $2.36\times10^{-2}$\tabularnewline
\hline 
 &  & $m_{2}$ & $1.25\times10^{-2}$ & $2.52\times10^{-2}$\tabularnewline
\hline 
 &  & $m_{3}$ & $5.03\times10^{-2}$ & $5.49\times10^{-2}$\tabularnewline
\hline 
 & 0.008 & $m_{1}$ & $7.76\times10^{-3}$ & $2.11\times10^{-2}$\tabularnewline
\hline 
 &  & $m_{2}$ & $1.17\times10^{-2}$ & $\mbox{2.28\ensuremath{\times}1\ensuremath{0^{-2}}}$\tabularnewline
\hline 
 &  & $m_{3}$ & $\mbox{\ensuremath{5.01\times10^{-2}}}$ & $5.38\times10^{-2}$\tabularnewline
\hline 
 & 0.007 & $m_{1}$ & $6.61\times10^{-3}$ & $1.86\times10^{-2}$\tabularnewline
\hline 
 &  & $m_{2}$ & $\mbox{1.09\ensuremath{\times}1\ensuremath{0^{-2}}}$ & $\mbox{2.05\ensuremath{\times}1\ensuremath{0^{-2}}}$\tabularnewline
\hline 
 &  & $m_{3}$ & $4.99\times10^{-2}$ & $5.29\times10^{-2}$\tabularnewline
\hline 
 & 0.006 & $m_{1}$ & $5.43\times10^{-3}$ & $1.61\times10^{-2}$\tabularnewline
\hline 
 &  & $m_{2}$ & $1.03\times10^{-2}$ & $1.83\times10^{-2}$\tabularnewline
\hline 
 &  & $m_{3}$ & $4.98\times10^{-2}$ & $5.20\times10^{-2}$\tabularnewline
\hline 
 & 0.005 & $m_{1}$ & $4.21\times10^{-3}$ & $1.36\times10^{-2}$\tabularnewline
\hline 
 &  & $m_{2}$ & $9.70\times10^{-3}$ & $1.62\times10^{-2}$\tabularnewline
\hline 
 &  & $m_{3}$ & $4.97\times10^{-2}$ & $5.13\times10^{-2}$\tabularnewline
\hline 
 & 0.004 & $m_{1}$ & $4.94\times10^{-3}$ & $1.13\times10^{-2}$\tabularnewline
\hline 
 &  & $m_{2}$ & $9.22\times10^{-3}$ & $1.42\times10^{-2}$\tabularnewline
\hline 
 &  & $m_{3}$ & $4.96\times10^{-2}$ & $5.08\times10^{-2}$\tabularnewline
\hline 
 & 0.003 & $m_{1}$ & $\mbox{1.61\ensuremath{\times}1\ensuremath{0^{-3}}}$ & $8.98\times10^{-2}$\tabularnewline
\hline 
 &  & $m_{2}$ & $8.89\times10^{-3}$ & $1.25\times10^{-2}$\tabularnewline
\hline 
 &  & $m_{3}$ & $4.95\times10^{-2}$ & $\mbox{5.03\ensuremath{\times}1\ensuremath{0^{-2}}}$\tabularnewline
\hline 
 & 0.002 & $m_{1}$ & $1.87\times10^{-4}$ & $6.84\times10^{-3}$\tabularnewline
\hline 
 &  & $m_{2}\mbox{}$ & $\mbox{8.74\ensuremath{\times}1\ensuremath{0^{-3}}}$ & $1.11\times10^{-2}$\tabularnewline
\hline 
 &  & $m_{3}$ & $4.95\times10^{-2}$ & $4.99\times10^{-2}$\tabularnewline
\hline 
 & 0.001 & $m_{1}$ & $1.35\times10^{-3}$ & $4.85\times10^{-2}$\tabularnewline
\hline 
 &  & $m_{2}$ & $\mbox{8.84\ensuremath{\times}1\ensuremath{0^{-3}}}$ & $9.99\times10^{-2}$\tabularnewline
\hline 
 &  & $m_{3}$ & $4.95\times10^{-2}$ & $4.97\times10^{-2}$\tabularnewline
\hline
\end{tabular}

\caption{Neutrino mass eigenvalue for normal hierarchy mass spectrum. Input
value are given in ref{[}21{]}}

\end{table}

\begin{table}[H]
\begin{tabular}{|c|c|c|c|c|}
\hline 
Mass Hierarchy & $m_{\nu}$ & Mass Eigenstate & Majorana Phases & Majorana Phases\tabularnewline
\hline
\hline 
 & (eV) & (eV) & $\phi_{2}=0^{0},\phi_{3}=180^{0}$ & $\phi_{3}=0^{0},\phi_{2}=180^{0}$\tabularnewline
\hline 
Inverted & 0.10 & $m_{1}$ & $1.04\times10^{-1}$ & $2.63\times10^{-2}$\tabularnewline
\hline 
 &  & $m_{2}$ & $1.04\times10^{-1}$ & $2.63\times10^{-2}$\tabularnewline
\hline 
 &  & $m_{3}$ & $9.18\times10^{-2}$ & $2.59\times10^{-2}$\tabularnewline
\hline 
 & 0.09 & $m_{1}$ & $9.33\times10^{-2}$ & $2.37\times10^{-2}$\tabularnewline
\hline 
 &  & $m_{2}$ & $9.38\times10^{-2}$ & $2.37\times10^{-2}$\tabularnewline
\hline 
 &  & $m_{3}$ & $7.98\times10^{-2}$ & $2.32\times10^{-2}$\tabularnewline
\hline 
 & 0.08 & $m_{1}$ & $8.29\times10^{-2}$ & $2.11\times10^{-2}$\tabularnewline
\hline 
 &  & $m_{2}$ & $8.33\times10^{-2}$ & $\mbox{2.11\ensuremath{\times}1\ensuremath{0^{-2}}}$\tabularnewline
\hline 
 &  & $m_{3}$ & $\mbox{\ensuremath{6.73\times10^{-2}}}$ & $2.05\times10^{-2}$\tabularnewline
\hline 
 & 0.07 & $m_{1}$ & $7.24\times10^{-2}$ & $1.84\times10^{-2}$\tabularnewline
\hline 
 &  & $m_{2}$ & $\mbox{7.29\ensuremath{\times}1\ensuremath{0^{-2}}}$ & $\mbox{1.85\ensuremath{\times}1\ensuremath{0^{-2}}}$\tabularnewline
\hline 
 &  & $m_{3}$ & $5.38\times10^{-2}$ & $1.78\times10^{-2}$\tabularnewline
\hline 
 & 0.06 & $m_{1}$ & $6.18\times10^{-2}$ & $1.58\times10^{-2}$\tabularnewline
\hline 
 &  & $m_{2}$ & $6.24\times10^{-2}$ & $1.58\times10^{-2}$\tabularnewline
\hline 
 &  & $m_{3}$ & $3.85\times10^{-2}$ & $1.51\times10^{-2}$\tabularnewline
\hline 
 & 0.05 & $m_{1}$ & $5.11\times10^{-2}$ & $1.32\times10^{-2}$\tabularnewline
\hline 
 &  & $m_{2}$ & $5.19\times10^{-2}$ & $1.32\times10^{-2}$\tabularnewline
\hline 
 &  & $m_{3}$ & $1.65\times10^{-2}$ & $1.23\times10^{-2}$\tabularnewline
\hline 
 & 0.04 & $m_{1}$ & $4.11\times10^{-2}$ & $1.06\times10^{-1}$\tabularnewline
\hline 
 &  & $m_{2}$ & $4.19\times10^{-2}$ & $1.06\times10^{-1}$\tabularnewline
\hline 
 &  & $m_{3}$ & $2.56\times10^{-2}$ & $9.43\times10^{-2}$\tabularnewline
\hline 
 & 0.03 & $m_{1}$ & $\mbox{3.09\ensuremath{\times}1\ensuremath{0^{-2}}}$ & $8.01\times10^{-2}$\tabularnewline
\hline 
 &  & $m_{2}$ & $3.22\times10^{-2}$ & $8.10\times10^{-2}$\tabularnewline
\hline 
 &  & $m_{3}$ & $3.71\times10^{-2}$ & $\mbox{6.38\ensuremath{\times}1\ensuremath{0^{-2}}}$\tabularnewline
\hline 
 & 0.02 & $m_{1}$ & $2.07\times10^{-2}$ & $5.46\times10^{-2}$\tabularnewline
\hline 
 &  & $m_{2}\mbox{}$ & $\mbox{2.25\ensuremath{\times}1\ensuremath{0^{-2}}}$ & $5.51\times10^{-2}$\tabularnewline
\hline 
 &  & $m_{3}$ & $4.38\times10^{-2}$ & $2.54\times10^{-2}$\tabularnewline
\hline 
 & 0.01 & $m_{1}$ & $1.01\times10^{-2}$ & $2.67\times10^{-2}$\tabularnewline
\hline 
 &  & $m_{2}$ & $\mbox{1.34\ensuremath{\times}1\ensuremath{0^{-2}}}$ & $2.81\times10^{-2}$\tabularnewline
\hline 
 &  & $m_{3}$ & $4.73\times10^{-2}$ & $4.03\times10^{-2}$\tabularnewline
\hline
\end{tabular}

\caption{Neutrino mass eigenvalue for inverted hierarchy mass spectrum. Input
value are given in ref{[}21{]}}

\end{table}

\begin{table}[H]
\begin{tabular}{|c|c|c|c|c|}
\hline 
Mass Hierarchy & $m_{\nu}$ & Mass Eigenstate & Majorana Phases & Majorana Phases\tabularnewline
\hline
\hline 
 & (eV) & (eV) & $\phi_{2}=0^{0},\phi_{3}=180^{0}$ & $\phi_{3}=0^{0},\phi_{2}=180^{0}$\tabularnewline
\hline 
Almost degenerate & 0.6 & $m_{1}$ & $1.58$ & $1.74$\tabularnewline
\hline 
 &  & $m_{2}$ & $1.58$ & 1.74\tabularnewline
\hline 
 &  & $m_{3}$ & $1.58$ & 1.74\tabularnewline
\hline 
 & 0.5 & $m_{1}$ & $1.32$ & 1.45\tabularnewline
\hline 
 &  & $m_{2}$ & $1.32$ & 1.45\tabularnewline
\hline 
 &  & $m_{3}$ & $1.32$ & 1.45\tabularnewline
\hline 
 & 0.4 & $m_{1}$ & $1.06$ & 1.16\tabularnewline
\hline 
 &  & $m_{2}$ & $1.06$ & 1.16\tabularnewline
\hline 
 &  & $m_{3}$ & $1.06$ & 1.16\tabularnewline
\hline 
 & 0.3 & $m_{1}$ & $0.79$ & 0.87\tabularnewline
\hline 
 &  & $m_{2}$ & $0.79$ & 0.87\tabularnewline
\hline 
 &  & $m_{3}$ & $0.79$ & 0.87\tabularnewline
\hline 
 & 0.2 & $m_{1}$ & $0.53$ & 0.58\tabularnewline
\hline 
 &  & $m_{2}$ & $0.53$ & 0.58\tabularnewline
\hline 
 &  & $m_{3}$ & $0.53$ & 0.58\tabularnewline
\hline 
 & 0.1 & $m_{3}$ & $0.26$ & 0.29\tabularnewline
\hline 
 &  & $m_{2}$ & $0.26$ & 0.29\tabularnewline
\hline 
 &  & $m_{3}$ & $0.26$ & 0.29\tabularnewline
\hline
\end{tabular}

\caption{Neutrino mass eigenvalue for degenerated mass spectrum. Input value
are given in ref{[}21{]}}

\end{table}

\section{Conclusions}

Future and present search for neutrinoless double beta decay purpose
at probing lepton number violation and the Majorona nature of neutrinos
with remarkable precession.Several experimental programs is currently
under discussion. We find the mass eigenvalue $m_{i}$in case of normal
hierarchy,inverted mass hierarchy and almost degenerate neutrino mass
spectrum. By taking $m_{ee}=(0.010-0.001)$ eV as reference value,
for normal mass hierarchy spectrum.We have predicted that $m_{1}$
varies from 0.0002eV - 0.01eV and $m_{2}$ varies from 0.009 eV- 0.013eV,
while variation in $m_{3}$ is from 0.049-0.056 eV. In case of inverted
hierarchy, $m_{1}$ and $m_{2}$ both vary from 0.01-0.263 eV while
$m_{3}$ varies from 0.047-0.259 eV for considered values of $\mbox{\ensuremath{m_{ee}}}$.
For 10-30 meV range of $m_{ee}$, $m_{3}$ values are higher than
$m_{1}$ and $m_{2}$, while for 40-100 meV range of $\mbox{\ensuremath{m_{ee}}}$,
$m_{3}$ value is lower than $m_{1}$ and $m_{2}$. The almost same
variation for $m_{1}$ and $m_{2}$ also confirms the consideration
of $m_{1}$, $m_{2}$, for inverted hierarchy. In case of AD hierarchy,
all the $m_{1}$, $m_{2}$, $m_{3}$ values lie  in the range from
0.26-1.74 eV. We have calculated the mass eigenvalue $m_{1},m_{2}$
and $m_{3}$ for all the three mass spectrum considered by taking
some specific choice of effective neutrino mass depending on the type
of mass spectrum. The ordering of mass states depends on choice of
$m_{ee}$, hence precise determination of $m_{ee}$ from single beta
decay experiment (Tritium beta decay) and future neutrino-less double
beta decay experiments will make the picture of mass spectrum clear.

\end{document}